# A new Rational Generating Function for the Frobenius Coin Problem

Deepak Ponvel Chermakani, IEEE Member

deepakc@pmail.ntu.edu.sg  d.chermakani@sms.ed.ac.uk  deepakc@usa.com  deepakc@myfastmail.com

**Abstract: -** An important question arising from the Frobenius Coin Problem is to decide whether or not a given monetary sum S can be obtained from N coin denominations. We develop a new Generating Function $G(x)$, where the coefficient of $x^i$ is equal to the number of ways in which coins from the given denominations can be arranged as a stack whose total monetary worth is $i$. We show that the Recurrence Relation for obtaining $G(x)$, is linear, enabling $G(x)$ to be expressed as a rational function, that is, $G(x) = P(x)/Q(x)$, where both $P(x)$ and $Q(x)$ are Polynomials whose degrees are bounded by the largest coin denomination.

## 1. Introduction

In the language of mathematics, the Frobenius Coin Problem is as follows: - Let $A_1, A_2, \ldots A_N$ be fixed positive integers. Find the largest integer S, such that $S \neq A_1 X_1 + A_2 X_2 + \ldots + A_N X_N$, for all non-negative integers $X_1, X_2, \ldots X_N$ [2]. An important question arising from the Frobenius Coin Problem is therefore to decide whether or not, for some given target positive integer S, there exist non-negative integers $X_1, X_2, \ldots X_N$, such that $S = A_1 X_1 + A_2 X_2 + \ldots + A_N X_N$ [1].

If we denote $F_i$ to be an integer sequence, such that $F_i$ is the number of ways in which i can be expressed using the given N coin denominations, then one can obtain a Generating Function given by SUMMATION ($F_i X^i$, i being integers $\geq 0$) [3]. The Generating Function helps predict whether or not some monetary sum S can be obtained using the given coin denominations, and is expressible as $P(x)/Q(x)$ where both $P(x)$ and $Q(x)$ are Polynomials. It has been proved that if N is fixed, then the Generating Function for $F_i$ can be expressed as a sum of short rational functions [4]. For $N \geq 4$, there is no explicit bound on the number of terms in $P(x)$ [3][5], though $Q(x)$ is bounded [3][5].

In this paper, we do not aim to develop a Generating Function explicitly for the Frobenius Coin Problem, i.e. for the sequence $F_i$ mentioned in the above paragraph. Instead, we develop a Generating Function for the sequence of integers (which we shall denote as $E_i$) representing the number of ways that coins from the given N denominations may be arranged as a stack, one coin over the other, such that the sum of the values of all coins in this stack add up to $i$. Clearly, the sequence $F_i$ is different from $E_i$, for example, given 2 coin denominations of 2 units and 5 units, then $F_7$ is 1, while $E_7$ is 2. It is also clear that $F_i$ is 0, if and only if, $E_i$ is 0. And we shall see that the Generating Function for $E_i$ can be expressed as a rational function, i.e. as $P(x)/Q(x)$, where both $P(x)$ and $Q(x)$ are Polynomials, and where the degrees, of both $P(x)$ and $Q(x)$, are equal to L, the largest coin denomination.

In the next Section 2, we shall prove a Theorem on a Linear Integer Recurrence Relation, and demonstrate how it represents the sequence $E_i$. In Section 3, we shall show the Generating Function for $E_i$.

## 2. The Linear Integer Sequence $E_i$

**Theorem-1:** Let L be the largest of the given N coin denominations, and the binary vector $<B_1, B_2, \ldots B_j, \ldots B_{L-1}, B_L>$ be such that $B_j$ is 1, if j is a coin denomination, and 0 otherwise. Then the number of ways that coins from the given denominations may be arranged as a stack, such that the total monetary worth of the stack is equal to $i$, is given by the following Linear Recurrence Relation for $E_i$

$E_0 = 1$, and,

$E_i = B_L E_{i-L} + B_{L-1} E_{i-L+1} + \ldots + B_2 E_{i-2} + B_1 E_{i-1}$ when $i$ is an integer greater than 0, and,

$E_i = 0$ when $i$ is an integer lesser than 0.

**Proof:** Expanding the sequence $E_i$ and expressing solely in terms of the B vector, we get the following

$E_1 = B_1$

$E_2 = B_2 + B_1^2$

$E_3 = B_3 + B_2 B_1 + B_1 B_2 + B_1^3 = B_3 + 2B_1 B_2 + B_1^3$

$E_4 = B_4 + B_3 B_1 + B_2^2 + B_2 B_1^2 + B_1 B_3 + 2B_1^2 B_2 + B_1^4 = B_4 + 2B_1 B_3 + B_2^2 + 3B_1^2 B_2 + B_1^4$

$E_5 = B_5 + 2B_1 B_4 + 2B_2 B_3 + 3B_2^2 B_1 + 4B_2 B_1^3 + 3B_1^2 B_3 + B_1^5$

and so on, showing that the sum of subscripts in each of the terms generated so far, are the same as the subscript of the corresponding $E_i$, and represent all possible stack formations. Next, let us prove by induction, that if we assume that the expressions via the B vector for $E_{i-L}$, $E_{i-L+1}$, … $E_{i-2}$, $E_{i-1}$ represent all possible stack formations with monetary worth equal to their corresponding subscripts, then $E_i = B_L E_{i-L} + B_{L-1} E_{i-L+1} + … + B_2 E_{i-2} + B_1 E_{i-1}$ will represent all possible stack formations of monetary value equal to $i$. To show this is true, we shall verify two things.

First, we verify that all terms of the form $B_1^{K_1} B_2^{K_2} B_3^{K_3} … B_L^{K_L}$ are represented in $E_i$ for non-negative integers $K_1$, $K_2 … K_L$ such that $K_1 + 2K_2 + … + LK_L = i$. If, for all integers $t$ less than $i$, one were to consider all possible ways of representing terms in $E_t$ (i.e. the sum of subscripts in the product should be equal to $i$), then it suffices to have the element $B_{i-t}$ appended to the terms of $E_t$, and represent that product in $E_i$, because every term in $E_p$ is the subset of some term in $E_q$, where $p < q$. For example, $E_t$ need not be multiplied by the product $B_{t1} B_{t2} B_{t3}$, where $t1+t2+t3 = i-t$, because terms generated by this product would already be coming into $E_i$ through other routes, namely from $B_{t1} E_{i-t1}$, from $B_{t2} E_{i-t2}$, and from $B_{t3} E_{i-t3}$.

Second, we verify the coefficient of the term $B_1^{K_1} B_2^{K_2} B_3^{K_3} … B_L^{K_L}$ in $E_i$, where $K_1, K_2 … K_L$ are positive integers. Clearly, this term gets linearly additive contributions from the coefficient of $B_1^{K_1-1} B_2^{K_2} B_3^{K_3} … B_L^{K_L}$ in $E_{i-1}$, from the coefficient of $B_1^{K_1} B_2^{K_2-1} B_3^{K_3} … B_L^{K_L}$ in $E_{i-2}$, …and so on, and finally from the coefficient of $B_1^{K_1} B_2^{K_2} B_3^{K_3} … B_L^{K_L-1}$ in $E_{i-L}$. So, if n! denotes the factorial of n, then the coefficient of $B_1^{K_1} B_2^{K_2} B_3^{K_3} … B_L^{K_L}$ in $E_i$, is equal to the following.

$$\frac{(K_1+K_2+…K_L-1)!}{(K_1-1)!\,K_2!\,K_3!…K_L!} + \frac{(K_1+K_2+…K_L-1)!}{K_1!\,(K_2-1)!\,K_3!…K_L!} + \frac{(K_1+K_2+…K_L-1)!}{K_1!\,K_2!\,(K_3-1)!…K_L!} + ….. + \frac{(K_1+K_2+…K_L-1)!}{K_1!\,K_2!\,K_3!…(K_L-1)!}$$

For each of the above L additive components, multiply and divide the $j^{th}$ component by $K_j$, to obtain the following.

$\frac{(K_1+K_2+…K_L)!}{K_1!\,K_2!\,K_3!…K_L!}$, which verifies to be the number of ways of arranging the term $B_1^{K_1} B_2^{K_2} B_3^{K_3} … B_L^{K_L}$ in a stack.

Finally, as $B_j$ is either 1 or 0, depending on whether or not coin denomination j is available, the value of $E_i$ will always equal the number of ways in which a monetary stack worth $i$, may be obtained from the given coin denominations.
**Hence Proved Theorem-1**

## 3. The Generating Function

The corresponding Generating Function G(x), for the sequence $E_i$ mentioned in Section 2, can be shown to be given by

$$G(x) = \frac{x^{L-1}(B_{L-1}+B_{L-2}E_1+…+B_1 E_{L-2}-E_{L-1}) + x^{L-2}(B_{L-2}+B_{L-3}E_1+…+B_1 E_{L-3}-E_{L-2})+…+x(B_1-E_1)-1}{B_L x^L + B_{L-1} x^{L-1} + … + B_1 x - 1}$$

## 4. Conclusion

In this paper, we presented a Linear Recurrence Relation for the integer sequence $E_i$, representing the number of ways in which a stack of coins worth $i$, may be built from the given coin denominations. This linearity enabled its Generating Function G(x) to be expressed as a rational function, i.e. as P(x)/Q(x), where both P(x) and Q(x) are Polynomials, and where the degrees of both P(x) and Q(x), are bounded by the largest coin denomination. As $E_i$ is zero, if and only if, a monetary amount of $i$ cannot be obtained from the coin denominations, G(x) will be of use to the Frobenius Coin Problem.

**About the Author**
I, Deepak Ponvel Chermakani, have written this paper, out of my own interest and initiative, during my spare time. I am currently a student at the University of Edinburgh UK (www.ed.ac.uk), where since Sep-2009, I have been enrolled in a fulltime one year Master Degree course in *Operations Research with Computational Optimization*. In Jul-2003, I completed a four year fulltime four year Bachelor Degree course in *Electrical and Electronic Engineering*, from Nanyang Technological University Singapore (www.ntu.edu.sg). I completed my high schooling from the National Public School in Bangalore in India in Jul-1999.